\documentclass[twocolumn,aps,prl,preprintnumbers]{revtex4}
\usepackage{}
\usepackage{bbm}
\usepackage{pifont}
\usepackage[latin9]{inputenc}
\usepackage{amsmath}
\usepackage{amssymb}
\usepackage{graphicx}
\usepackage{epstopdf}
\usepackage{pifont}
\usepackage{mathrsfs}
\usepackage{amsfonts}
\usepackage{amsthm}
\usepackage{color}
\usepackage{txfonts}
\usepackage[colorlinks=true,citecolor=blue,linkcolor=blue,urlcolor=blue,anchorcolor=blue]{hyperref}%
\hypersetup{colorlinks=true,citecolor=blue,linkcolor=blue,urlcolor=blue}
\providecommand{\U}[1]{\protect\rule{.1in}{.1in}}
\makeatletter
\@ifundefined{textcolor}{}
{
\definecolor{BLACK}{gray}{0}
\definecolor{WHITE}{gray}{1}
\definecolor{RED}{rgb}{1,0,0}
\definecolor{GREEN}{rgb}{0,1,0}
\definecolor{BLUE}{rgb}{0,0,1}
\definecolor{CYAN}{cmyk}{1,0,0,0}
\definecolor{MAGENTA}{cmyk}{0,1,0,0}
\definecolor{YELLOW}{cmyk}{0,0,1,0}
}
\makeatother
\begin{document}
\title{Observation of type-III corner states induced by long-range interactions}
\author{Huanhuan Yang}
\author{Lingling Song}
\author{Yunshan Cao}
\author{Peng Yan}
\email[Corresponding author: ]{yan@uestc.edu.cn}
\affiliation{School of Electronic Science and Engineering and State Key Laboratory of Electronic Thin Films and Integrated Devices, University of Electronic Science and Technology of China, Chengdu 610054, China}

\begin{abstract}
Long-range interactions (LRIs) are ubiquitous in nature. Higher-order topological (HOT) insulator represents a new phase of matter. A critical issue is how LRI dictates the HOT phases. In this work, we discover four topologically distinct phases, i.e., HOT phase, the bound state in the continuum, Dirac semimetal (DSM) phase, and trivial insulator phase in a breathing kagome circuit with tunable LRIs. We find an emerging type-III corner state in the HOT phase, which splits from the edge state continuum and originates from the strong couplings between nodes at different edges. We experimentally detect this novel state by impedance and voltage measurements. In the DSM phase, the Dirac cone exhibits an itinerant feature with a tunable position that depends on the LRI strength. Our findings provide a deeper understanding of the LRI effect on exotic topological states and pave the way for regulating interactions in topolectrical circuits.
\end{abstract}

\maketitle
\section{introduction}
Long-range interactions (LRIs) typically represent the two-body potential decaying algebraically at large distances with a power smaller than the spatial dimension. Paradigms of LRIs include electromagnetism, gravity, 2D vortices, etc. Physical systems with LRIs can exhibit many peculiar properties in their dynamics and statistics, such as negative specific heat and temperature jumps \cite{Campa2009,French2010}. Recently, the LRIs have been shown to impose certain effects on the topological states \cite{Varney2010,Beugeling2012,Li2019,Shen2021,Olekhno2022,Yu2017,Leykam2018}. For instance, Li \emph{et al.} observed a type-II corner state in the photonic breathing kagome crystal in the presence of a weak LRI \cite{Li2019}. An open question is how a strong enough LRI dictates the topological phase. Quantum-optical technology may be a choice to address this issue with controllable spin-spin interactions, but it is difficult to implement in experiments \cite{Sandvik2010,Britton2012}.

Recently, electrical circuit manifests as an exceptional platform to study topological physics \cite{Olekhno2022,Jia2015,Albert2015,Imhof2018,Ezawa2018,Lee2018,Hadad2018,Hofmann2019,
JBao2019,WZhang2020,Helbig2020,Liu2020,
Olekhno2020,Song2020,Yang2020,SLiu2020,RChen2020,Zhang2020,Yang2021,Yangyt2021,Yang2022,Ventra2022,Wu2022,
NAO2022,Galeano2022,Pei2022,Zhang2022}, which can imitate the tight-binding (TB) model in condensed matter physics by the inductor-capacitor ($LC$) network. Conveniently, one can introduce an arbitrary hopping term between different nodes in the circuit. For example, the higher-order topological (HOT) insulators have been widely explored in circuits \cite{Imhof2018,Ezawa2018,JBao2019,WZhang2020,Song2020,Yang2020,SLiu2020,RChen2020}, which however focused on short-range interactions or hoppings.

\begin{figure*}[htbp!]
  \centering
  % Requires \usepackage{graphicx}
  \includegraphics[width=0.98\textwidth]{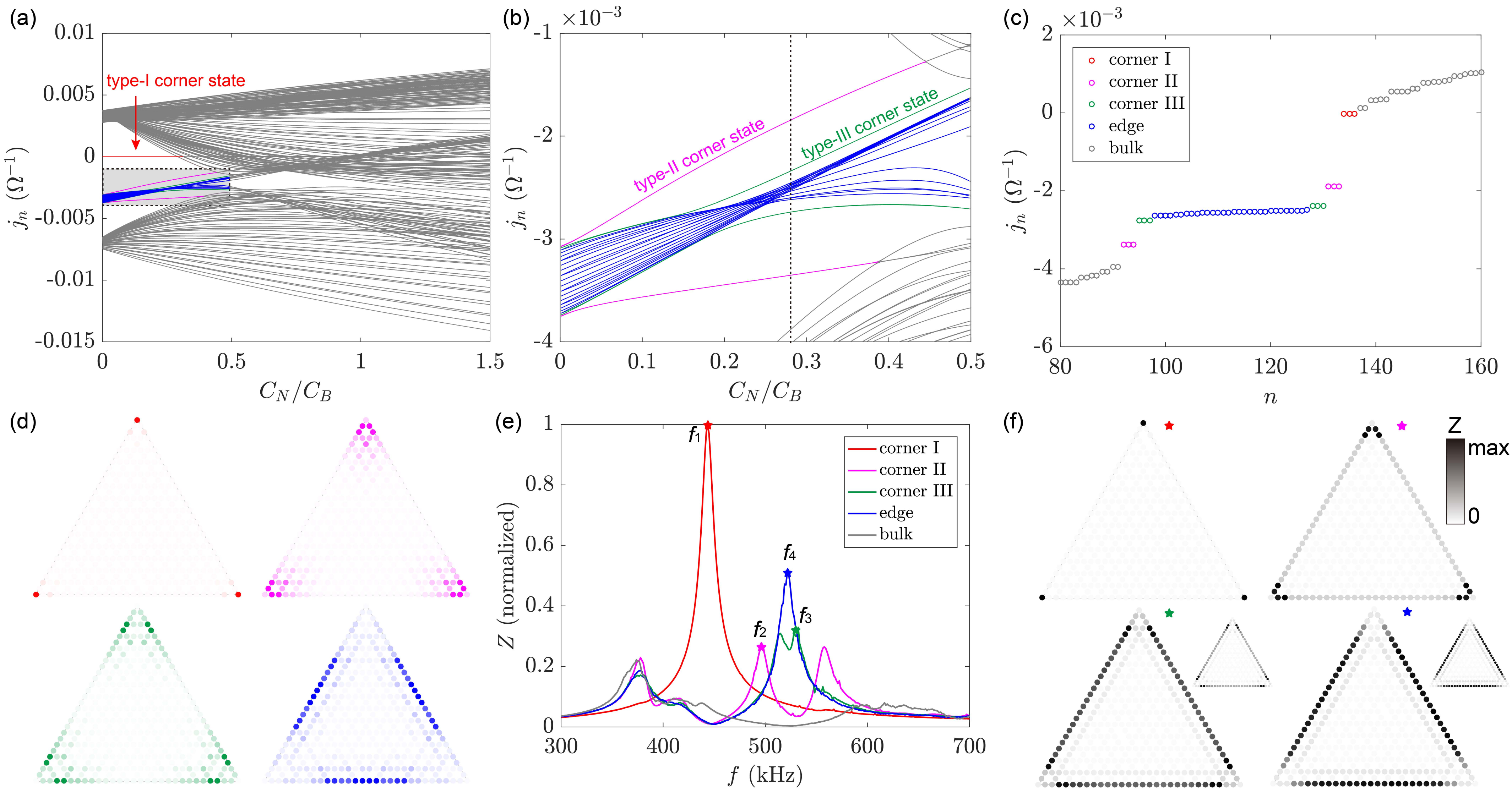}\\
  \caption{Admittance spectrums, wave functions, and impedances. (a) The admittance spectrum as a function of the capacitance ratio $C_N/C_B$ at $\omega=\omega_c$. The red, magenta, green, blue, and gray curves represent the type-I corner, type-II corner, type-III corner, edge, and bulk states, respectively. (b) The details of the admittance spectrum [gray rectangle in (a)]. (c) Admittance spectrum at $C_N/C_B=0.275$ [dashed black line in (b)]. (d) The wave-function profiles of the type-I corner, type-II corner, type-III corner, and edge states. (e) The impedance of the 1st, 3rd, 9th, 84th, and 78th node as a function of the driving frequencies. (f) The distributions of impedance at different frequencies. Colored pentagrams on the top right corner indicate the measured frequencies marked in (e). Insets:  The distributions of impedance of type-III corner and edge state with higher $Q$-factor inductors ($Q=200$).}\label{EIG}
\end{figure*}

It is thus interesting to establish a fruitful connection between these two different topical areas of physics, i.e., LRIs and HOT phases. In this work, we investigate how LRIs affect the HOT insulators both in theory and experiment. To this end, we first derive the full phase diagram which includes four topologically distinct phases in the breathing kagome circuit with LRIs, specifically, a HOT phase, a bound state in the continuum (BIC, higher-order topology supporting corner-localized bound states in the continuum \cite{WABenalcazar2020}), a Dirac semimetal (DSM) phase, and a trivial insulator phase. Interestingly, the HOT phase can be further classified into three types of corner states. The type-I corner states are pinned to zero admittance due to the protection of the generalized chiral symmetry, and they are localized at the three vertexes of the lattice. The type-II and type-III corner states are at finite admittances and split from the edge state continuum, which are akin to the edge states but exponentially decaying away from the 2nd and 3rd cells along the boundary, respectively [see the 2nd and 3rd subfigures in Fig. \ref{EIG}(d)]. Moreover, the type-II corner states separate from the lowest and highest edge modes and appear for an arbitrary LRI. By contrast, the type-III ones split from the edge state continuum only in the strong-coupling region, namely, a critical LRI value is required to move them out of the edge band. By measuring the impedance distribution and the voltage signal propagation, we directly observe all three types of corner states in experiments. In addition, we numerically identify the BIC and DSM phases. We show that the position of the Dirac points depends on the strength of LRI, which may help to control the transport behaviors of Dirac states.

The paper is organized as follows. In Sec. \ref{SecII}, we present the circuit model and derive the full phase diagram. The emerging corner states are analyzed in Sec. \ref{SecIII}. We perform experimental measurements in Sec. \ref{SecIV}. Conclusions are drawn in Sec. \ref{SecV}. Technical details are given in Appendixes.
\begin{figure*}[htbp!]
  \centering
  % Requires \usepackage{graphicx}
  \includegraphics[width=0.98\textwidth]{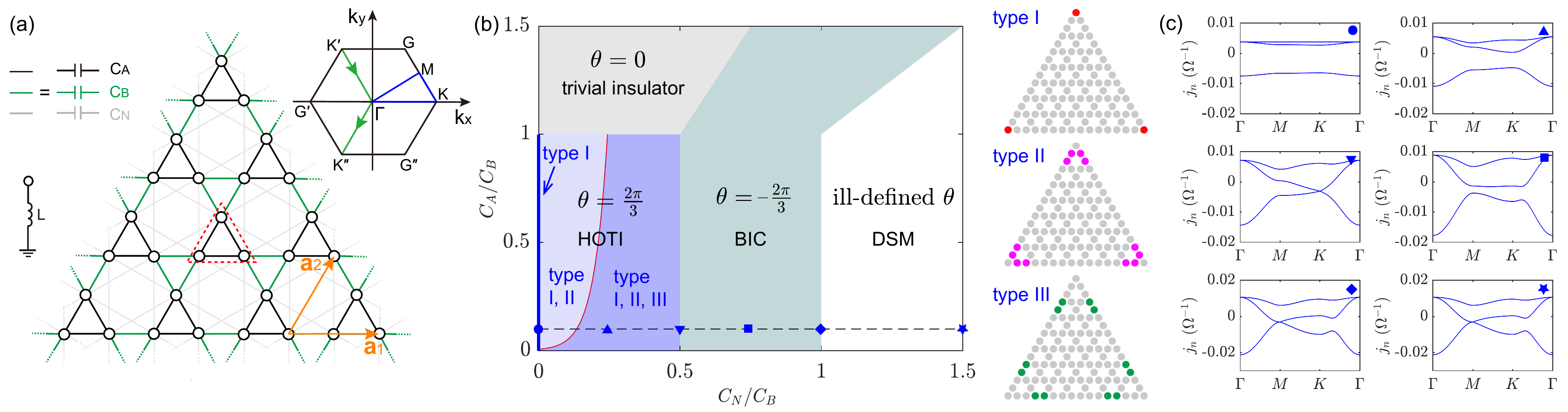}\\
  \caption{Breathing kagome circuit with LRIs supporting four topologically distinct phases. (a) Schematic plot of an infinite breathing kagome circuit with NNN hoppings. The breathing means alternatively arranged intracell and intercell hoppings $C_A$ and $C_B$. The black, green, and gray line segments represent the capacitors $C_A$, $C_B$, and $C_N$, respectively, and each node is grounded by an inductor $L$. The basic vectors ${\bf a}_1=(1,0)$ and ${\bf a}_2=(1/2,\sqrt{3}/2)$. Inset: the first BZ. (b) Phase diagram of $\mathbb{Z}_3$ Berry phase by tuning the ratio $C_A/C_B$ and $C_N/C_B$. The blue, green, white, and gray regions represent the HOT phase, BIC phase, DSM phase, and trivial insulator phase, respectively. The red curve indicates the critical NNN hopping strength for the emergence of type-III corner states. Insets: schematic plots of three types of corner states. (c) At $C_A/C_B=0.1$ ($C_B=1$ nF), the band structures for $C_N/C_B=0$, 0.25, 0.5, 0.75, 1, and 1.5, respectively, corresponding to the blue symbols (from left to right) in (b).}\label{model_inf}
\end{figure*}
\begin{figure*}[htbp!]
  \centering
  % Requires \usepackage{graphicx}
  \includegraphics[width=0.98\textwidth]{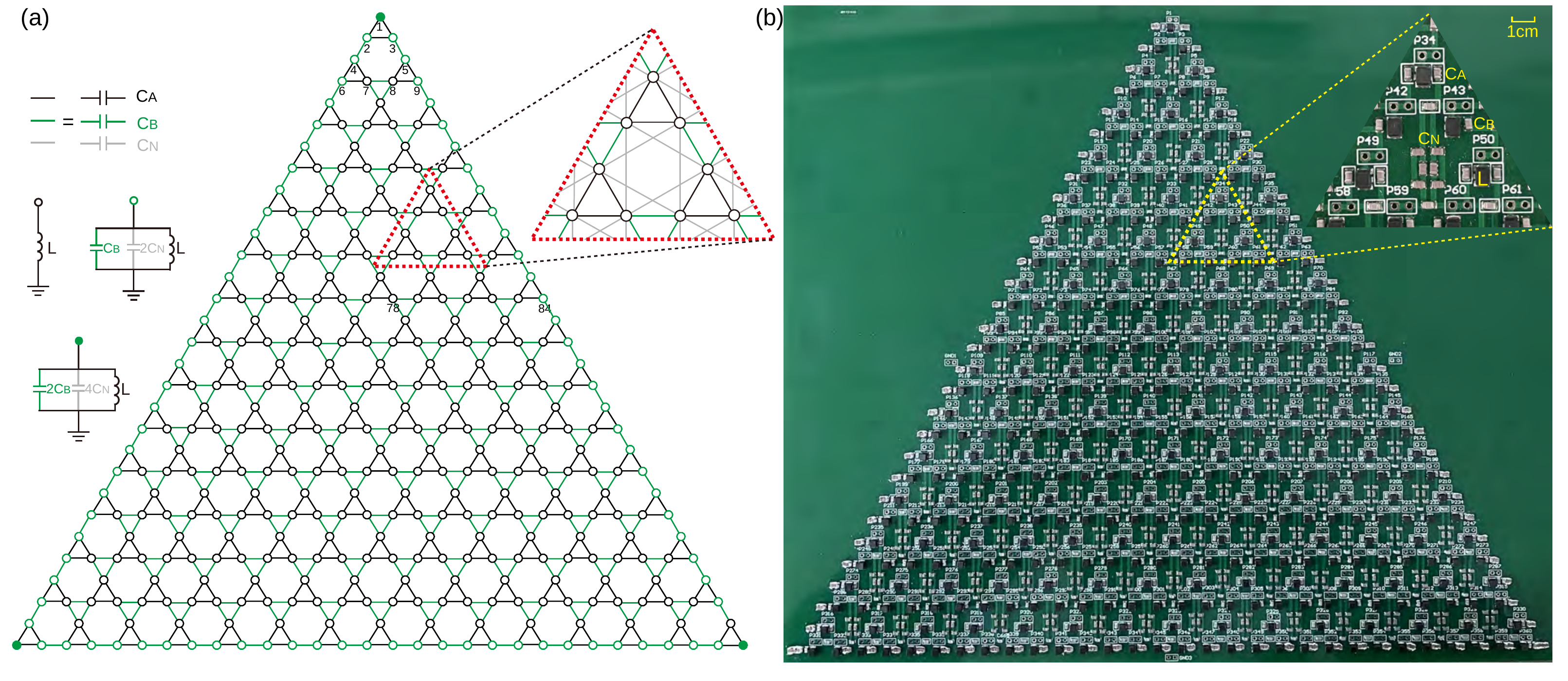}\\
  \caption{Finite-size circuit model and the photograph of experimental setup. (a) Illustration of a finite-size kagome circuit with 360 nodes. The gray, green, and gray line segments represent the capacitors $C_A$, $C_B$, and $C_N$, respectively, and each node is grounded by capacitors and inductors with configurations shown in the insets. (b) Photograph of the printed circuit broad used in experiments. The inset zooms in the local details of the circuit.}\label{model}
\end{figure*}

\section{Circuit model and phase diagram}\label{SecII}
As shown in Fig. \ref{model_inf}(a), we consider an infinite breathing kagome $LC$ circuit with LRIs, by introducing next-nearest-neighbor (NNN) hopping terms (capacitors $C_N$) to the lattice. The NNN interactions can be regarded as LRIs for two reasons. In most topological systems with true LRIs, one can consider the NNN hopping terms without missing the key physics, because higher-order terms beyond NNN can only contribute negligible effects \cite{Li2019,Shen2021}. Besides, the concept of LRI very recently has been adopted in systems with NNN interactions \cite{Olekhno2022}. Labelling the nodes of the circuit by $a=1,2,...$ the response at frequency $\omega$ follows Kirchhoff's law: $I_a(\omega)=\sum_bJ_{ab}(\omega)V_b(\omega)$, with $I_a$ the external current flowing into node $a$, $V_b$ the voltage at node $b$, and $J_{ab}(\omega)$ being the circuit Laplacian: $J_{ab}(\omega)=i\omega\left[-C_{ab}+\delta_{ab}\left(\sum_nC_{an}-\frac{1}{\omega^2L_a}\right)\right]
$ with $C_{ab}$ the capacitance between nodes $a$ and $b$, and $L_a$ being the grounded inductance at node $a$. The sum is taken over all connected nodes. The dashed red triangle represents the unit cell including three nodes, with the first Brillouin zone (BZ) plotted in the inset of Fig. \ref{model_inf}(a). For convenience, we express $J(\omega)$ as $i\mathcal{H}(\omega)$, with $\mathcal{H}(\omega)$ akin to the Hermitian TB Hamiltonian expressed as
\begin{equation}\label{H}
 \mathcal {H}=\mathcal {H}_0+\mathcal {H}_\text{NNN},
\end{equation}
where $\mathcal{H}_0$  and $\mathcal{H}_{\rm NNN}$ represent the NN and NNN coupling Hamiltonian, respectively (see Appendix \ref{A} for details). Next, as suggested in Refs. \cite{Hatsugai2011,Kariyado2018,Kudo2019,Araki2020}, we use $\mathbb{Z}_3$ Berry phase to distinguish these phases in the parameter space (here is the $C_A$-$C_N$ plane), which is computed by
\begin{equation}\label{Z3}
 \mathcal{\theta}=\int_{L_{1}}\text{Tr}[\textbf{A}(\textbf{k})]\cdot d\textbf{k}\ \  (\text{mod}\ 2\pi),
\end{equation}
where $\textbf{A}(\textbf{k})$ is the Berry connection: $\textbf{A}(\textbf{k})=i\Psi^{\dag}(\textbf{k})\frac{\partial}{\partial\textbf{k}}\Psi(\textbf{k})$
with $\Psi(\textbf{k})$ being the eigenvector of \eqref{H} for the lowest band. $L_{1}$ is an integral path in BZ: $K^{\prime}\rightarrow \Gamma\rightarrow K^{\prime\prime}$, shown by the green line segment in the inset of Fig. \ref{model_inf}(a) (see Appendix \ref{A} for discussion). The computation of $\mathbb{Z}_3$ Berry phase is displayed in Fig. \ref{model_inf}(b). Here, we obtain three different $\mathbb{Z}_3$ Berry phases, i.e., $\mathbb{Z}_3=2\pi/3$, $-2\pi/3$, $0$, and an ill-defined value, filled by blue, green, gray, and white colors, respectively. Specifically, $\mathbb{Z}_3=2\pi/3$ denotes the HOT phase, $\mathbb{Z}_3=-2\pi/3$ indicates the BIC phase (see Appendix \ref{B}), and $\mathbb{Z}_3=0$ is the trivial insulator phase. The ill-defined $\mathbb{Z}_3$ belongs to the DSM phase owing to the gapless band structures ($\mathbb{Z}_N$ Berry phase can only be rigorously defined for a gapped system. In the presence of strong NNN couplings, the lowest two bands touch each other. One thus cannot have a well-defined $\mathbb{Z}_3$). It is noted that, in the absence of the NNN coupling, one can only observe the type-I corner state and trivial phase by tuning the ratio $C_A/C_B$ \cite{Yang2020}. However, by adjusting certain capacitances of $C_A$ or $C_B$ in this lattice by breaking the crystal symmetry (such as $C_3$ symmetry), one may expect some peculiar edge states as illustrated in Ref. \cite{Yang2022}.

Figures \ref{model_inf}(c) displays the band structures for different capacitance ratio $C_N/C_B$ at $C_A/C_B=0.1$ [blue symbols in Fig. \ref{model_inf}(b)].  For $0<C_A/C_B<1$, in the absence of NNN hopping term ($C_N=0$), the band structures are gapped, which belongs to the HOT phase \cite{Yang2020}. As we increase the capacitance $C_N$, the lowest two bands firstly converge at $K$ point at $C_N/C_B=0.5$ and reopen beyond it, and then close again at $M$ point when $C_N/C_B =1$  (see Appendix \ref{A}). The degenerate point subsequently moves along the $M\rightarrow\Gamma$ trace, forming the DSM state (see Appendix \ref{C}). This phenomenon may provide a feasible method to manage the unique transport properties of itinerant Dirac states, in exploring the physics of anomalous quantum Hall effect \cite{Zhang2005}, minimum conductivity \cite{Ando2002,Tworzydlo2006}, and Klein tunneling \cite{Katsnelson2006,Stander2009}. For $C_A/C_B >1$, the two gapless thresholds appear at $C_A/C_B =2C_N/C_B$  and  $C_A/C_B =C_N/C_B$, and the behaviors of band structures are similar to the foregoing ones (not shown). Below, we focus on the HOT phase, i.e., $0<C_A/C_B <1$ and $0<C_N/C_B <0.5$ [blue region in Fig. \ref{model_inf}(b)].
\begin{figure*}[htbp!]
  \centering
  % Requires \usepackage{graphicx}
  \includegraphics[width=0.98\textwidth]{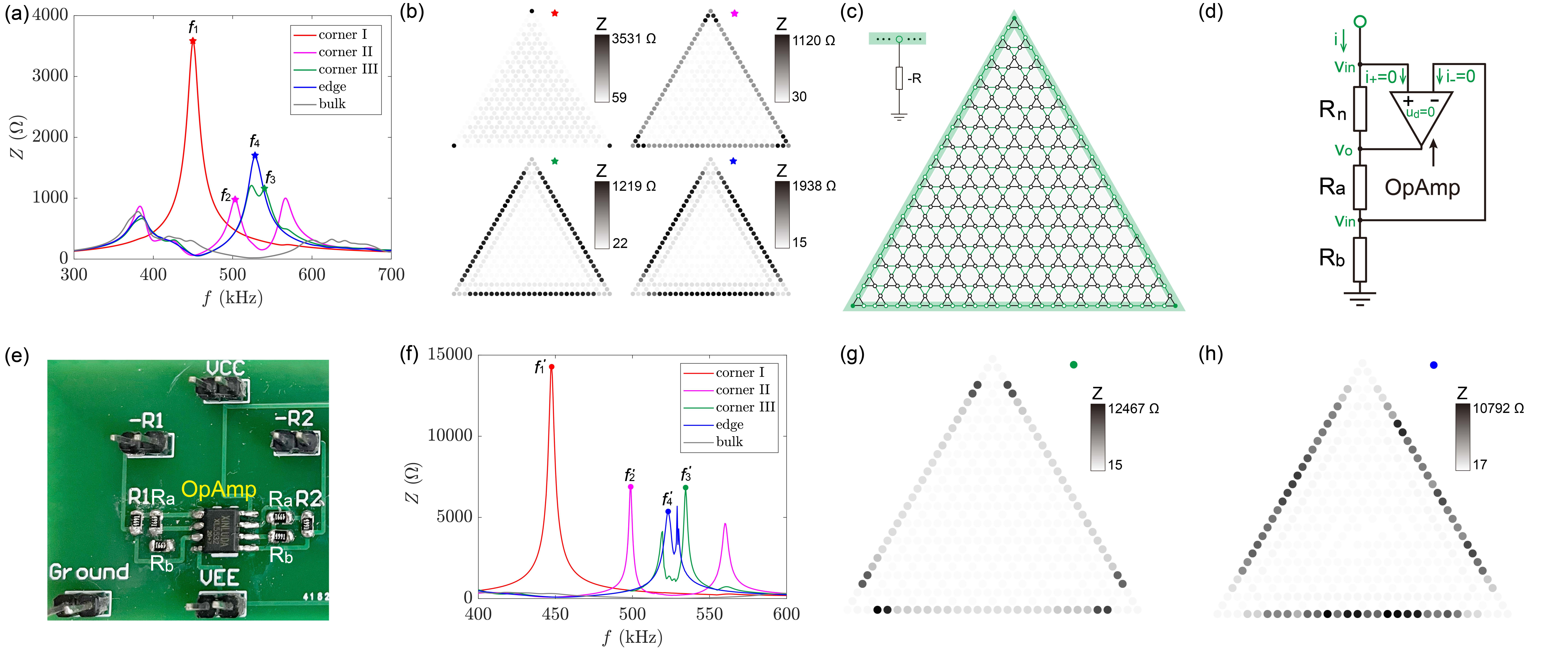}\\
  \caption{Experimental measurements of the impedance. (a) Experimental impedance dependence on the driving frequency. The red, magenta, green, and blue pentagrams mark the positions of the impedance peak of type-I corner, type-II corner, type-III corner, and edge states ($f_1=450$ kHz, $f_2=503$ kHz, $f_3=537$ kHz, and $f_4=528$ kHz). (b) The distributions of experimental impedance at the four frequencies $f_1$, $f_2$, $f_3$, and $f_4$. (c) The negative resistance is connected to each boundary node inside the green shadow. The schematic diagram (d) and the experimental realization (e) of the negative resistance (here, we adopt $R_n=R_a=R_b=4.99$ k$\Omega$). (f) Experimental impedance as a function of the driving frequency with negative resistances, with $f_1'=447$ kHz, $f_2'=499$ kHz, $f_3'=534$ kHz, and $f_4'=523$ kHz. The impedance distribution of (g) the type-III corner state and (h) edge state with negative resistances measured at frequency $f_3'$ and $f_4'$, respectively.}\label{Exp}
\end{figure*}

\section{Corner states}\label{SecIII}
To observe the corner states, we consider a finite-size circuit network with $\mathcal{N}=360$ nodes with the configuration being shown in Fig. \ref{model}(a). Diagonalizing the circuit Laplacian $J(\omega)$, we obtain both the admittance spectrum and wave functions. In the following calculations, we consider $C_A/C_B =0.1$ with $C_B=1$ nF. Figure \ref{EIG}(a) shows the admittance spectrum for different NNN hopping strengths. Except for the isolated zero-admittance mode (red segment), the other two modes (magenta and green segments) escape from the edge spectrums, which represent the type-II and type-III corner states, see Fig. \ref{EIG}(b) for details. We demonstrate that the type-II corner states appear for any $C_N>0$, while the type-III corner states emerge only when $C_N>C_{Nc}$ with $C_{Nc}$ the critical value, as shown by the dark blue region in Fig. \ref{model_inf}(a) (see Appendix \ref{D} for details). In Ref. \cite{Li2019}, the same lattice is constructed in a photonic system but with a rather weak LRI. Therefore, only the type-II corner states were observed there, and they split from both the lowest and highest boundary modes. In order to find the type-III corner states, one needs a stronger LRI, and the circuit then exhibits its superiority for such a purpose. The strong NNN interactions induce strong couplings between nodes at different edges, so the internal edge modes can separate from the edge state continuum and form the type-III corner states.

Next, we analyze a representative example at $C_N/C_B =0.275$ [the dashed black line in \ref{EIG}(b)]. Specifically, we choose $C_N=0.275$ nF and $L=39$ $\mu$H. The admittances $j_n$ are depicted in Fig. \ref{EIG}(c), with five kinds of eigenmodes donated by red, magenta, green, blue, and gray circles, corresponding to type-I corner, type-II corner, type-III corner, edge, and bulk states, respectively. It is noted that the type-I corner states are pinned to zero admittance, while type-II and type-III corner states are at finite admittances. Figure \ref{EIG}(d) shows the wave functions of the corner states and edge states, from which one can distinguish them clearly. The wave functions of the type-II (type-III) corner states are resembling those of the edge states, which have been proved to be topological \cite{Ni2019,Li2019}, but exponentially decaying away from the 2nd (3rd) cell along the boundary. This behavior can be viewed as the boundary states of the topological edge modes, i.e., the corner states.

In an electrical circuit, one can use the impedance between the node and ground to reflect the mode near the zero admittance \cite{Lee2018,Yang2020,Yang2021}. In order to observe the different states mentioned above, one can shift the corresponding modes to zero admittance by changing the driving frequency without modifying the wave functions. Here, we adopt the impedance between the 1st, 3rd, 9th, 84th, 78th nodes and the ground to quantify the signals from the type-I corner, type-II corner, type-III corner, edge, and bulk states [see numbered nodes in Fig. \ref{model}(a)], respectively. Figure \ref{EIG}(e) shows the theoretical impedance as a function of the driving frequency, in which the red, magenta, green, blue, and gray curves denote the signal from the aforementioned five nodes, respectively. It is noted that we consider the quality factor ($Q$ factor) of inductor here with the value $Q=45$ measured in the experiment. One can see that the impedance peaks of the type-I, type-II, type-III, and edge states appear at $f_1=444$ kHz, $f_2=497$ kHz, $f_3=530$ kHz, and $f_4=520$ kHz, respectively, marked by four colored pentagrams, which means the four states close to zero admittance at these frequencies.

We calculate the distributions of the impedance at $f_1, f_2, f_3$ and $f_4$, with the results plotted in Fig. \ref{EIG}(f). At frequency $f_1=444$ kHz, the type-I corner modes are pinned at zero admittance. We depict the distribution of impedance at the resonant frequency in the first subfigure of Fig. \ref{EIG}(f). The impedances concentrate on the three vertexes of the sample, which confirms the existence of the type-I corner states. For type-II corner states, we find the distributions of the impedance are identical to the corresponding eigenmode profiles at frequency $f_2$. However, the impedance curve of the edge state behaves as an extension and submerges the type-III corner states due to the finite $Q$ factor of inductor. Therefore, the distributions of the impedance for type-III corner states seem to sprinkle throughout the edge. To distinguish the type-III corner states from edge states, one needs to improve the $Q$ factor of inductors. We plot the distributions of impedance with higher $Q$-factor inductors ($Q=200$) in the insets of Fig. \ref{EIG}(f), then one can distinguish the type-III corner states from the edge states. Next, we verify these numerical results by experiments.

\section{Experiment}\label{SecIV}
As shown in Fig. \ref{model}(b), we manufacture a printed circuit broad (PCB) to observe three kinds of corner states. To directly compare with the numerical calculations, we choose the same electric elements as those in theoretical analysis, but each circuit element has a natural 2\% tolerance. The impedance of circuit is measured by the impedance analyzer (Keysight E4990A). Figure \ref{Exp}(a) shows the experimental impedances at different driving frequencies. Similarly, we adopt the impedances between the 1st, 3rd, 9th, 84th, 78th nodes and the ground to characterize the type-I corner (red curve), type-II corner (magenta curve), type-III corner (green curve), edge (blue curve), and bulk (gray curve) states, respectively. The experimental results are fully consistent with the numerical ones. Then, we measure the distribution of impedance at four frequencies marked by red, magenta, green, and blue pentagrams in Fig. \ref{Exp}(a), with the results plotted in Fig. \ref{Exp}(b). One can observe the type-I corner states, type-II corner states clearly, but the type-III corner states mix with the edge states due to the low $Q$ factor.

The $Q$ factor of an inductor is defined as $Q=\omega L/r$, where the $r$ comes from inevitable losses. To compensate for the dissipation, we introduce active elements to the boundary nodes [light green region in Fig. \ref{Exp}(c)], realized by connecting a series of negative resistances $-R$ with the configuration shown in Figs. \ref{Exp}(d) and \ref{Exp}(e) (see Appendix \ref{E}). We numerically compare two cases that $-R$ in introduced to all nodes and to boundary nodes only, which gives almost identical results. We therefore only add negative resistances to edge nodes to reduce the experimental complexity. As a consequence, the effective $Q$-factor of inductors is improved to about 150, so the impendence peak becomes very sharp, with the peak value being enhanced by about 400\%, as shown in Fig. \ref{Exp}(f). We then measure the distribution of impedance at frequency $f_3'$ and $f_4'$ again and clearly separate the type-III corner states from the edge states, see Figs. \ref{Exp}(g) and \ref{Exp}(h). From these experimental results, we find that all corner states are robust against the disorder because the elements intrinsically have 2\% tolerance.

\begin{figure}
  \centering
  % Requires \usepackage{graphicx}
  \includegraphics[width=0.48\textwidth]{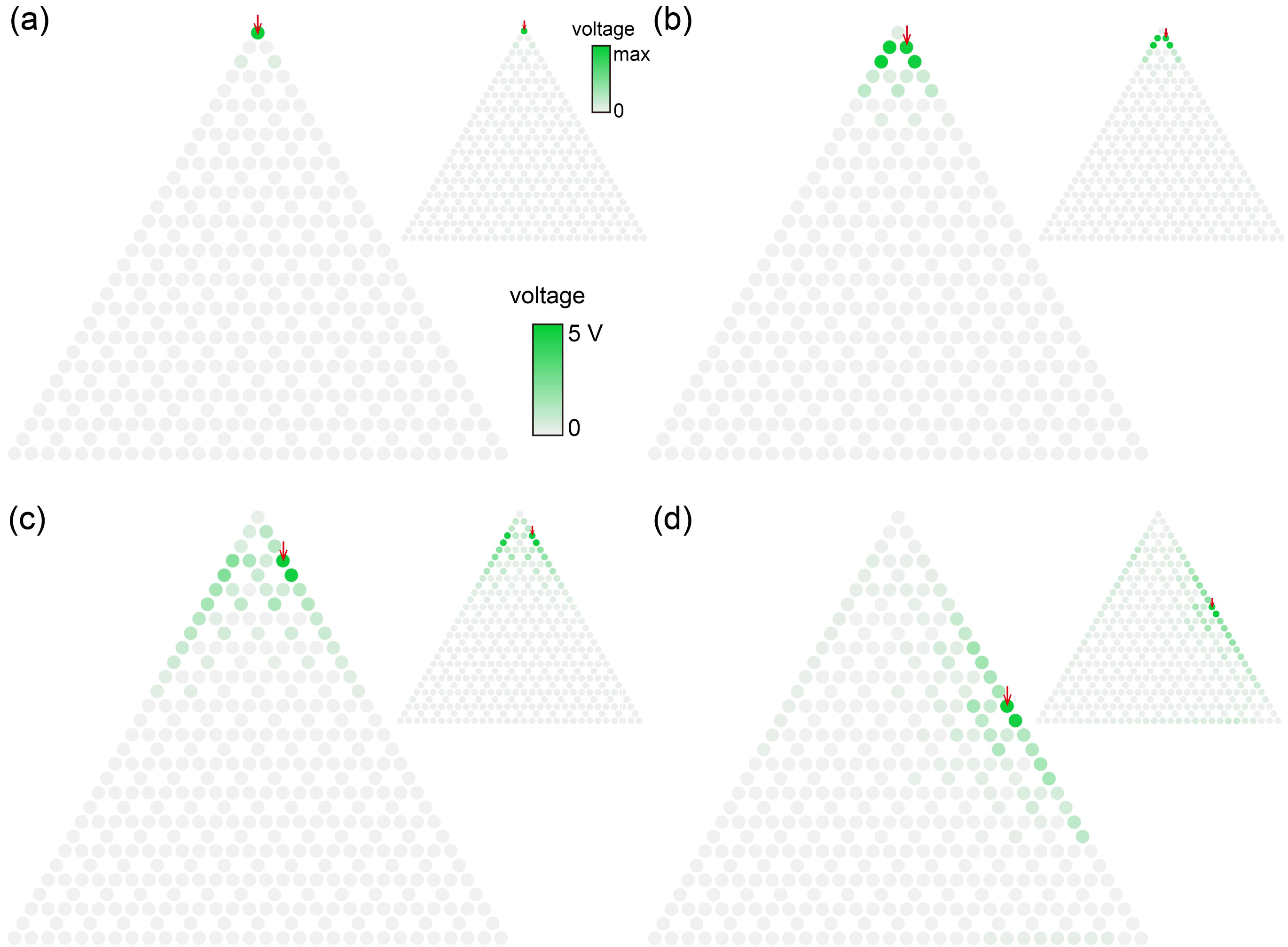}\\
  \caption{Experimental observation and theoretical calculation of the voltage propagations. (a)-(d) The distributions of the amplitude of the voltage signal in devices with the signal source imposed between the 1st, 3rd, 9th, 84th node and the ground at frequency $f_1'=447$ kHz, $f_2'=499$ kHz, $f_3'=534$ kHz, and $f_4'=523$ kHz, respectively. Insets: numerical results. }\label{current}
\end{figure}

To directly observe the dynamic response of the circuit, we measure the propagation of the voltage signals for all three corner states and edge states. As shown in Fig. \ref{current}(a), we impose a voltage signal at the first node (labeled by red arrow) by $v(t)=v_0 \sin (2\pi ft)$ with the amplitude $v_0=5$ V and $f=f_1'$ by an arbitrary function generator (GW AFG-3022), and measure the amplitudes of the steady voltage sign by the oscilloscope (Keysight MSOX3024A) over the whole sample. We find that the signal is only localized at the first node, indicating the type-I corner state. With the same method, we futher observe the type-II corner, type-III corner, and edge states at frequency at the $f_2'$, $f_3'$, and $f_4'$, with the results plotted in Figs. \ref{current}(b)-\ref{current}(d), respectively. Meanwhile, we show theoretical results in the insets of each subfigure for comparison, which agree well with experimental findings.

\section{Conclusion}\label{SecV}
To summarize, we investigated the effect of LRIs on the topological phases in breathing kagome circuits. With appropriate LRIs, we discovered the type-III corner state and presented a direct experimental observation. We showed that the type-III corner states split from the edge modes, and originate from the strong coupling between nodes at different edges. With stronger LRIs, the BIC and DSM phases subsequently emerge in this system. Our results highlighted the important role played by LRI in topological phases and phase transitions. Our findings significantly advance the understanding of the localization behavior of HOT states and may spur future studies in other systems, such as acoustic lattices, photonic crystals, and cold atoms, and in higher dimensions allowing peculiar hinge states and Weyl semimetal states. It is worth mentioning that the NNN hopping terms in our system differ from the LRIs in the field of strongly correlated electrons that are closely related to local and non-local density interactions. LRIs beyond long-range hopping may exhibit other appealing properties, which deserve thorough exploration in the future.

\begin{acknowledgments}
\section*{ACKNOWLEDGEMENTS}
This work was supported by the National Natural Science Foundation of China (Grants No. 12074057, No. 11604041, and No. 11704060).
\end{acknowledgments}
\section*{APPENDIX}
\begin{appendix}
\section{The generalized chiral symmetry, $\mathbb{Z}_3$ Berry phase and phase transitions}\label{A}
\begin{figure}
  \centering
  % Requires \usepackage{graphicx}
  \includegraphics[width=0.35\textwidth]{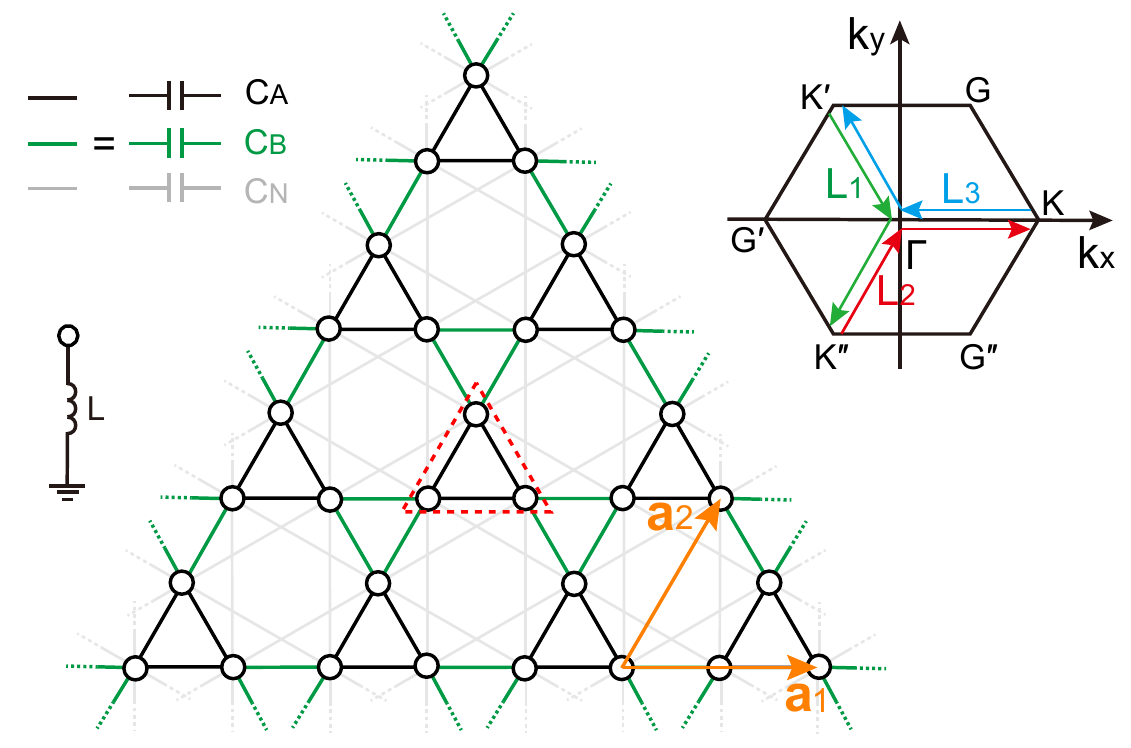}\\
  \caption{Schematic plot of an infinite-size kagome circuit composed of three types of capacitors $C_A$, $C_B$, and $C_N$, and each node is grounded by an inductor. Inset: the first BZ with three colored arrows indicating the integral paths.}\label{model2}
\end{figure}

As shown in Fig. \ref{model2}, the NN coupling Hamiltonian is given by
\begin{equation}\label{H0}
 \mathcal {H}_0=\left(
 \begin{matrix}
   Q_{0} & Q_{1} & Q_{2}\\
   Q_{1}^{*} & Q_{0} & Q_{3} \\
   Q_{2}^{*} & Q_{3}^{*} & Q_{0}
  \end{matrix}
  \right),
\end{equation}
with the matrix elements
\begin{equation} \label{Q0}
\begin{aligned}
Q_{0}&=\omega\left[2(C_{A}+C_{B})-\frac{1}{\omega^{2}L}\right],\\
Q_{1}&=-\omega[C_{A}+C_{B}e^{-i(k_{x}/2+\sqrt{3}k_{y}/2)}],\\
Q_{2}&=-\omega[C_{A}+C_{B}e^{-ik_{x}}],\\
Q_{3}&=-\omega[C_{A}+C_{B}e^{-i(k_{x}/2-\sqrt{3}k_{y}/2)}],
\end{aligned}
\end{equation}
where $k_{x(y)}$ is the wave vector in the $\hat{x}(\hat{y})$ directions, and
\begin{equation}\label{HNNN}
 \mathcal {H}_\text{NNN}=\left(
 \begin{matrix}
   Q_{N0} & Q_{N1} & Q_{N2}\\
   Q_{N1}^{*} & Q_{N0} & Q_{N3} \\
   Q_{N2}^{*} & Q_{N3}^{*} & Q_{N0}
  \end{matrix}
  \right),
\end{equation}
with the matrix elements
\begin{equation} \label{QN0}
\begin{aligned}
Q_{N0}&=4\omega C_N,\\
Q_{N1}&=-\omega C_N[e^{-ik_{x}}+e^{i(k_{x}/2-\sqrt{3}k_{y}/2)}],\\
Q_{N2}&=-\omega C_N[e^{-i(k_{x}/2-\sqrt{3}k_{y}/2)}+e^{-i(k_{x}/2+\sqrt{3}k_{y}/2)}],\\
Q_{N3}&=-\omega C_N[e^{-ik_{x}}+e^{i(k_{x}/2+\sqrt{3}k_{y}/2)}].
\end{aligned}
\end{equation}

It can be verified that $\mathcal{H}_0$ satisfies the generalized chiral symmetry at resonant frequency $\omega_{0}=1/[\sqrt{ L(2C_A+2C_B)}]$, because of $\Gamma_{3}^{-1}\mathcal {H}_0\Gamma_{3}=\mathcal {H}_{1}$,
 $\Gamma_{3}^{-1}\mathcal {H}_{1}\Gamma_{3}=\mathcal {H}_{2}$,
 $\mathcal {H}_0+\mathcal {H}_{1}+\mathcal {H}_{2}=0$,
with $\Gamma_{3}=\text{diag}\left(1,e^{\frac{2\pi i}{3}},e^{\frac{4\pi i}{3}}\right)$.
If we add the diagonal element $Q_{N0}$ to $Q_0$, the resonant frequency will shift from $\omega_0$ to $\omega_{c}=1/[\sqrt{ L(2C_A+2C_B+4C_N)}]$. We find that the remainder of $\mathcal {H}_\text{NNN}$ still holds the generalized chiral symmetry. It is to say that the periodic NNN hopping terms do not break the generalized chiral symmetry but only cause a shift of the resonant point, which provides topological protection for the triply degenerate zero-admittance mode.

Next, we discuss the relationship between the $C_3$ symmetry and the quantized $\mathbb{Z}_3$ Berry phase. The Berry phase is closely related to the local twist of the Hamiltonian \cite{Hatsugai2011}, which is quantized due to time-reversal symmetry and inversion symmetry, for instance. In the breathing kagome lattice, it has $C_3$ symmetry, so the points $K$, $K'$, $K''$ in the Brillouin zone are equivalent, as shown in the inset of Fig. \ref{model2}. So, there are three equivalent paths for computing $\mathbb{Z}_3$ Berry phase: $L_1$: $K^{\prime}\rightarrow \Gamma\rightarrow K^{\prime\prime}$, $L_2$: $K^{\prime\prime}\rightarrow \Gamma\rightarrow K$ and $L_3$: $K\rightarrow \Gamma\rightarrow K^{\prime}$. As a result, the Berry phases $\theta(L_i)$ should take the same value, i.e., $\theta(L_1)=\theta(L_2)=\theta(L_3)$. In addition, the integral along the path $L_1+L_2+L_3$ is equal to zero, i.e., $\sum_{i=1,2,3}\theta(L_i)=0$. Therefore, we obtain the quantized $\mathbb{Z}_3$ Berry phase $\theta \equiv \theta(L_i)=2\pi k/3$ with $k=0,1,2$.

To show the details of phase transition at $K$ and $M$ points, we analyze the eigenvalues and wave functions at these two points. At $K$ point, the admittance spectra are given by $j_K^{(1)}=C_A+C_B-4C_N$, $j_K^{(2)}=C_A-2C_B+2C_N$, and $j_K^{(3)}=-2C_A+C_B+2C_N$ with the corresponding wave functions expressed as $\phi_1=\left[-\frac{\sqrt{3}}{2}i-\frac{1}{2},\frac{\sqrt{3}}{2}i-\frac{1}{2},1\right]^T$,
$\phi_2=\left[\frac{\sqrt{3}}{2}i-\frac{1}{2},-\frac{\sqrt{3}}{2}i-\frac{1}{2},1\right]^T$
and $\phi_3=[1,1,1]^T$. We find that the lowest band exchanges from $j_K^{(2)}$ to $j_K^{(1)}$ at $C_N/C_B=0.5$ [the first phase transition in Fig. \ref{model_inf}(b)]. Meanwhile, the expectation value of the phase of wave functions for the lowest band $\varphi_{C_3}={\rm arg}(\phi^\dagger R_3\phi$) \cite{Ni2019} changes from $\frac{2\pi}{3}$ to $-\frac{2\pi}{3}$ (same as the $\mathbb{Z}_3$ Berry phase), where $R_3=
\left(
  \begin{array}{ccc}
    0 & 0 & 1 \\
    1 & 0 & 0 \\
    0 & 1 & 0 \\
  \end{array}
\right)
$ is the three-fold rotational operator.

At $M$ point, the admittance spectra are $j_M^{(1)}=C_A+C_B-2C_N$, $j_M^{(2)}=C_N-C_B/2-C_A/2-\Delta/2$, and $j_M^{(3)}=C_N-C_B/2-C_A/2+\Delta/2$ with the eigenmodes $\phi_1=[0,-1,1]^T$, $\phi_2=\left[-(C_A+C_B-2C_N-\Delta)/(2C_A-2C_B),1,1\right]^T$, and
$\phi_3=\left[-(C_A+C_B-2C_N+\Delta)/(2C_A-2C_B),1,1\right]^T$. Here $\Delta=\sqrt{9C_A^2-14C_AC_B-4C_AC_N+9C_N^2-4C_BC_N+4C_N^2}$. The lowest two bands close at $C_N/C_B=1$, accompanied by the exchange of the lowest band from $j_M^{(2)}$ to $j_M^{(1)}$ and the wave functions from $\phi_2$ to $\phi_1$ [the second phase transition in Fig. \ref{model_inf}(b)]. In the case of $C_N/C_B>1$, the lowest two bands touch each other and form the semimetal states. For the trivial state, the wave-function phase $\varphi_{C_3}$ is always equal to 0.

\section{Bound state in the continuum}\label{B}
\begin{figure}[htbp!]
  \centering
  % Requires \usepackage{graphicx}
  \includegraphics[width=0.48\textwidth]{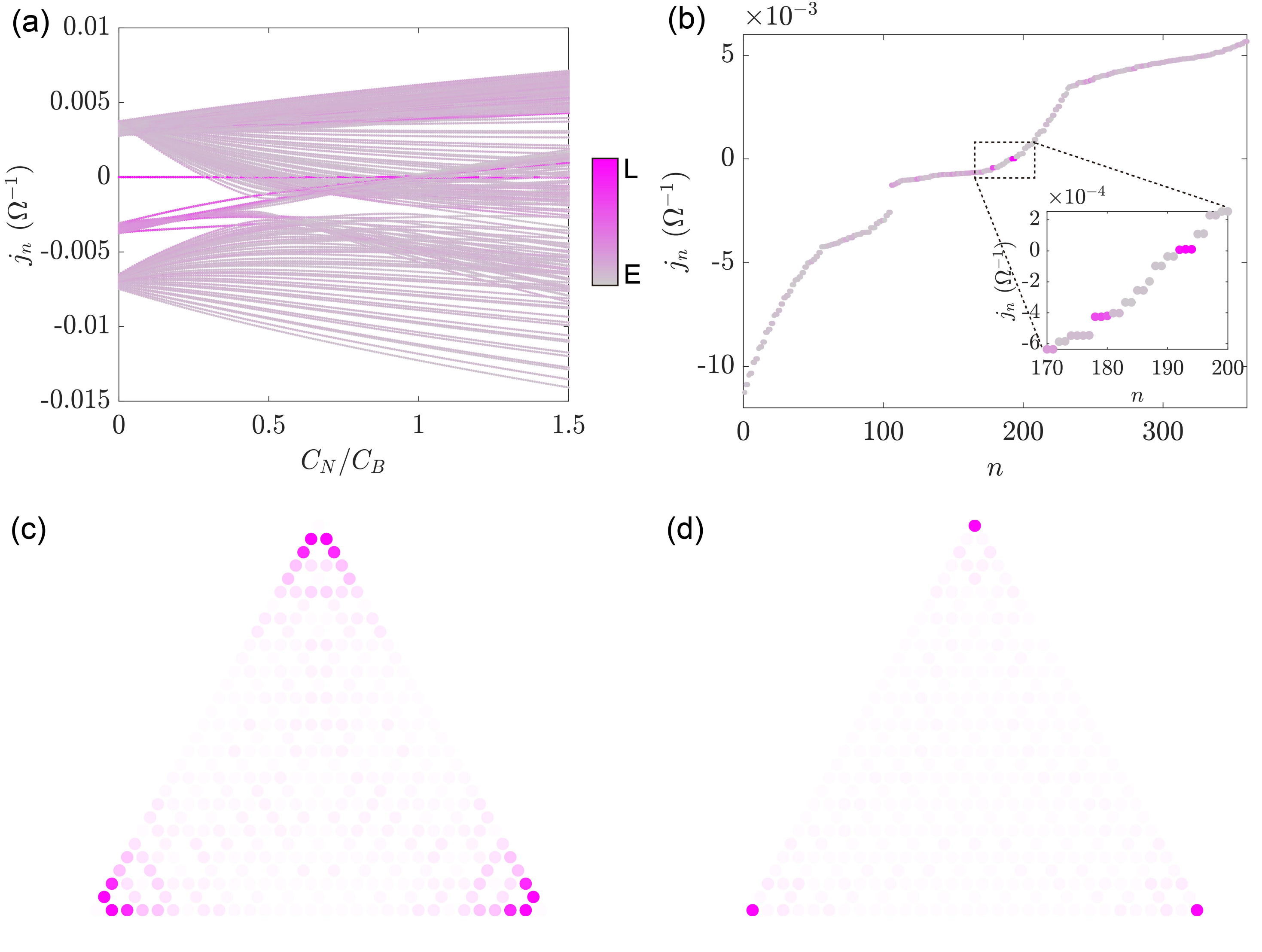}\\
  \caption{Admittance spectrums and eigenmodes of BIC. (a) The admittance spectrum as a function of $C_N/C_B$ at $C_A/C_B=0.1$, with the IPR indicated by the color map. The magenta dots represent the localized states, while gray dots mean extended states. (b) The admittance spectrum at $C_N/C_B=0.75$ with the inset showing the details. (c)(d) The profile of the wave functions of the localized states with the state number $n=180$ and 194, respectively.}\label{local}
\end{figure}

In Fig. \ref{EIG}(a), one cannot find any isolated mode in the range of $0.5<C_N/C_B <1$ [green region in Fig. \ref{model_inf}(b)]. However, if we evaluate the inverse participation ratio (IPR)
\begin{equation}
p=\sum_i|\phi_n|^4
\end{equation}
of the system, where $\phi_n$ is the normalized wave function with $\sum_i|\phi_n|^2=1$, we find localized states buried in the bulk and edge continuum, as shown in Fig. \ref{local}(a). The IPR has been widely adopted to study Anderson localization in disordered systems \cite{Janssen1998}, and to confirm the existence of HOT states recently \cite{HAraki2019,Wakao2020}. Taking $C_N/C_B=0.75$ as an example, we find two groups of localized states appearing in the continuum, as shown in Fig. \ref{local}(b). We plot their wave functions in Figs. \ref{local}(c) and \ref{local}(d), respectively. We confirm that both localized states actually are corner states.

\section{Dirac semimetal state}\label{C}
\begin{figure}
  \centering
  % Requires \usepackage{graphicx}
  \includegraphics[width=0.48\textwidth]{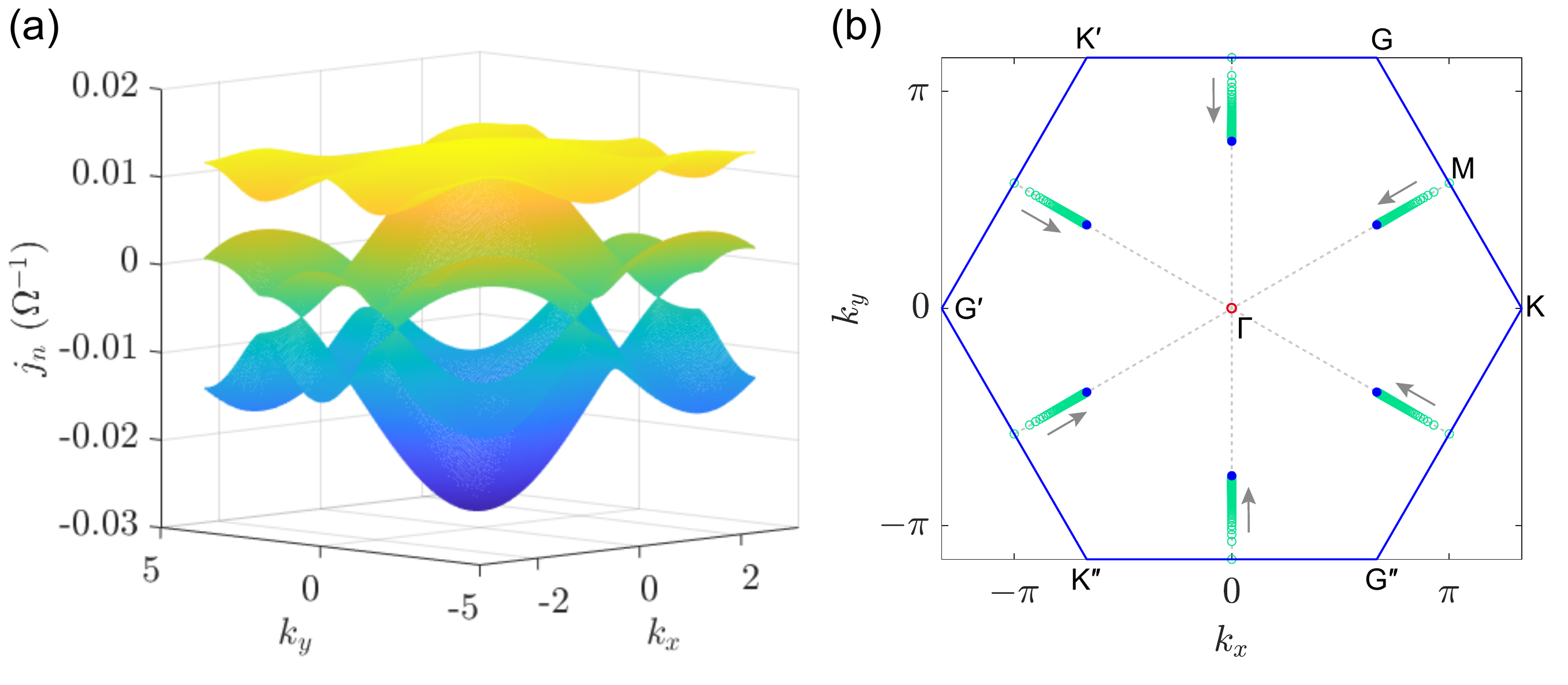}\\
  \caption{Tunable Dirac points. (a) Illustration of the band structures of an infinite system with $C_A/C_B=0.1$ and $C_A/C_N=1.5$. (b) The positions of the six Dirac points (green dots) move in the first BZ as $C_N/C_B$ ranges from 1 to 20. The blue hexagon shows the first BZ and the gray arrows indicate the moving direction of the Dirac points.}\label{metal}
\end{figure}

To illustrate the DSM state, we plot the band structure of an infinite kagome circuit in Fig. \ref{metal}(a). The lowest two bands linearly touch each other at six points, which corresponds to the DSM state. Besides, by tuning the capacitance $C_N$, we find that the positions of the six Dirac points continuously move (one of them moves along the path $M\rightarrow\Gamma$), see Fig. \ref{metal}(b).

To obtain an analytical understanding, we begin with Hamiltonian \eqref{H}, and rewrite it as a concise form
\begin{equation}
\mathcal{H}=\left(
              \begin{array}{ccc}
                0 & a & b \\
                a^* & 0 & c \\
                b^* & c^* & 0 \\
              \end{array}
            \right),
\end{equation}
with $a=Q_1+Q_{N1}$, $b=Q_2+Q_{N2}$, and $c=Q_3+Q_{N3}$ [see Eq. \eqref{Q0} and Eq. \eqref{QN0} for details]. The secular equation is given by
\begin{equation}
\lambda^3-(aa^*+bb^*+cc^*)\lambda-ab^*c-a^*bc^*=0.
\end{equation}

For a cubic equation, if the discriminant $27(ab^*c+a^*bc^*)^2=4(aa^*+bb^*+cc^*)^3$ is satisfied, it must have three real roots with two degenerate ones. Imposing $k_y=\frac{\sqrt{3}}{3}k_x$, the matrix elements $a,b,$ and $c$ can be simplified as
\begin{equation}
\begin{aligned}
a&=b=-\omega(C_A+C_Be^{-ik_x}+C_Ne^{-ik_x}),\\
c&=-\omega(C_A+2C_N\cos k_x).
\end{aligned}
\end{equation}
Divided by $-\omega C_B$, the above equations are expressed as
\begin{equation}
\begin{aligned}
a&=b=t_a+e^{-ik_x}+t_ne^{-ik_x},\\
c&=t_a+2t_n\cos k_x,
\end{aligned}
\end{equation}
with $t_a=C_A/C_B$ and $t_n=C_N/C_B$. The discriminant is simplified as $8(aa^*)^3-15(aa^*)^2c^2+6aa^*c^4+c^6=0$. Substituting the expressions of $a,c$ to this equation, we get
\begin{equation}
 2 t_a(t_n-1) \cos k_x + 2 t_n^2 \cos(2k_x)+ t_n^2 -2t_n -1=0.
\end{equation}
The solution is given by
\begin{equation} \label{DT}
\cos(k_x)=-\frac{t_a - t_at_n + \sqrt{t_a^2t_n^2 - 2t_a^2t_n + t_a^2 + 4t_n^4 + 8t_n^3 + 4t_n^2}}{4t_n^2}.
\end{equation}

At $t_n=1$, we have $\cos(k_x)=-1$ ($k_x=\pm \pi$). In the case of $t_n<1$ ($t_n>1$), the value of the right part of Eq. \eqref{DT} is greater (less) than 1, so we can obtain a real solution of $k_x$ only for $t_n>1$. In the limit of $t_n\rightarrow\infty$, we obtain $\lim_{t_n\rightarrow\infty}\cos{k_x}=-1/2$ and $k_x=\pm\frac{2\pi}{3}$, which is the terminal of the Dirac points. We therefore prove that the solution $k_y=\frac{\sqrt{3}}{3} k_x$ is indeed the trajectory of the Dirac points. Due to the $C_3$ symmetry of the lattice, the other two equivalent traces of Dirac points exist along the lines of $k_y=-\frac{\sqrt{3}}{3}k_x$ and $k_x=0$, respectively. To show the linear touching of lowest two bands, we consider the circumstance of $C_N\rightarrow\infty$ with Dirac points at $(k_x^{\rm DP},k_y^{\rm DP})=(\pm\frac{2\pi}{3},\pm\frac{2\sqrt{3}\pi}{9})$ and $(0,\pm\frac{4\sqrt{3}\pi}{9})$, marked by the blue dots in Fig. \ref{metal}(b). In such a case, we can analytically solve the band structure as
\begin{equation}
\begin{aligned}
&j^{(1)}_n=2\omega C_N,\\
&j^{(2)}_n=\omega C_N\left[-1-\sqrt{3+4\cos\frac{3k_x}{2}\cos\frac{\sqrt{3}k_y}{2}
+2\cos(\sqrt{3}k_y)}\right],\\
&j^{(3)}_n=\omega C_N\left[-1+\sqrt{3+4\cos\frac{3k_x}{2}\cos\frac{\sqrt{3}k_y}{2}
+2\cos(\sqrt{3}k_y)}\right].
\end{aligned}
\end{equation}

In the vicinity of the Dirac points, one can expand the lowest two bands [$j^{(2),(3)}_n$] as
\begin{equation}
j_n(p_x,p_y)=\omega C_N[-1+\frac{3\sqrt{2}}{2}p_x+3p_y+\mathscr{O}({\bf p^2})],
\end{equation}
with $p_x=k_x-k_x^{\rm DP}$, $p_y=k_y-k_y^{\rm DP}$ and $|{\bf p}|\ll1$. One can clearly see a linear band crossings near the Dirac points.

\section{Origin of the type-II and type-III corner states}\label{D}

\begin{figure}
  \centering
  % Requires \usepackage{graphicx}
  \includegraphics[width=0.48\textwidth]{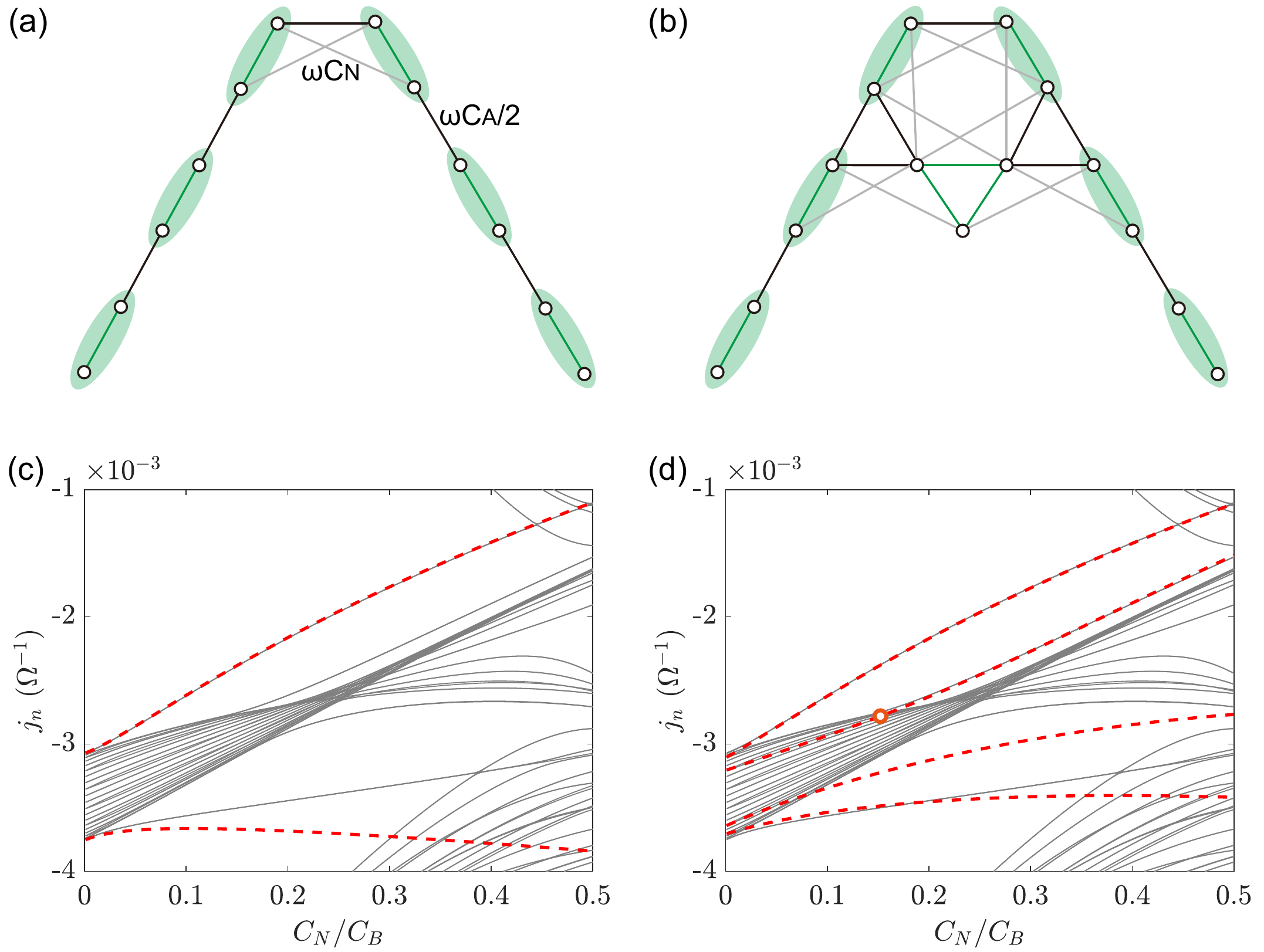}\\
  \caption{Simplified models explaining the origin of type-II and type-III corner states. (a)(b) The first-order and second-order perturbation models. (c)(d) The numerical admittance spectrums (gray lines) and the perturbed ones (dashed red lines).}\label{per}
\end{figure}

To explain the origin of the type-II corner states, we consider the first-order perturbation theory as shown in Fig. \ref{per}(a). The perturbation Hamiltonian is given by
\begin{equation} \label{Eper}
V=-\omega\left(
    \begin{array}{cccccc}
      0 &  \frac{C_A}{2} & 0 & 0 & 0 & 0 \\
       \frac{C_A}{2} & 0 &  \frac{C_A}{2} & 0 & 0 & 0 \\
      0 & \frac{C_A}{2} & 0 &  \frac{C_A}{2}+C_N & 0 & 0 \\
      0 & 0 &  \frac{C_A}{2}+C_N & 0 &  \frac{C_A}{2} & 0 \\
      0 & 0 & 0 &  \frac{C_A}{2} & 0 &  \frac{C_A}{2} \\
      0 & 0 & 0 & 0 & \frac{C_A}{2} & 0 \\
    \end{array}
  \right).
\end{equation}
Diagonalizing the above matrix, we obtain the perturbation solution
\begin{equation}
\delta j_n=\pm\omega\frac{(C_A+2C_N)^2+C_A^2}{2C_A+4C_N},
\end{equation}
with the results plotted by dashed red lines in Fig. \ref{per}(c). The two solutions are type-II corner states. The NNN couplings cause an effective interaction ($\omega C_N$) between two dimers across the corner, which splits the type-II corner states from the edge spectrums.

For type-III corner states, we have to consider the second-order perturbation theory as shown in Fig. \ref{per}(b). The system is described by a $15\times15$ Hamiltonian matrix $V'$ (not shown). By diagonalizing it, we obtain the dashed red spectrums in Fig. \ref{per}(d), which can explain the type-III corner states well. The NNN couplings induce extra interactions to edge nodes by connecting to the bulk nodes, which yields the type-III corner states.

\begin{figure}
  \centering
  % Requires \usepackage{graphicx}
  \includegraphics[width=0.5\textwidth]{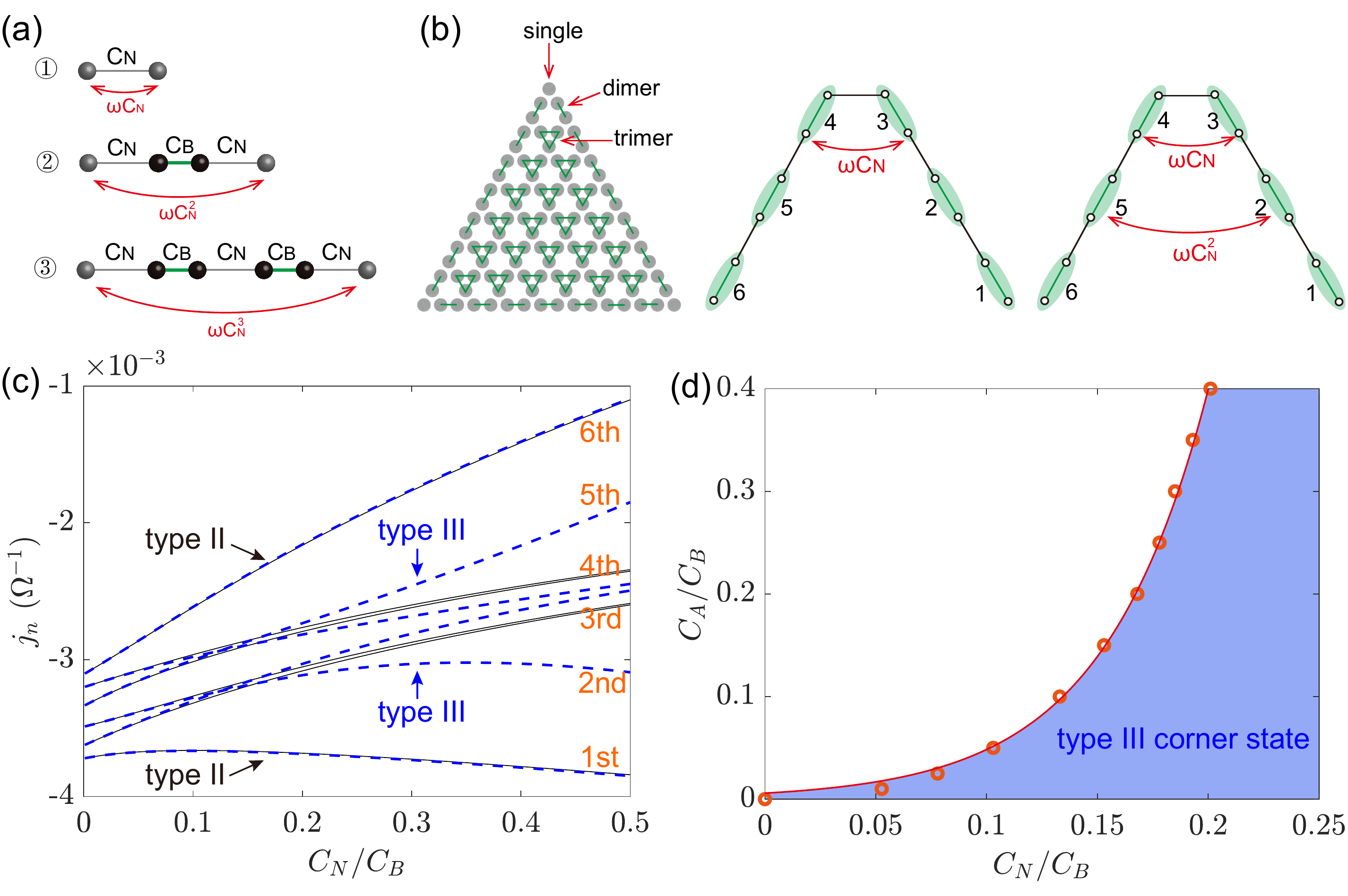}\\
  \caption{Effective interactions between edge nodes. (a) SSH model with different bulk nodes. The red arrows represent the effective capacitance for energy splitting. (b) Effective interactions between edge nodes. (c) The admittance spectrum of the two effective models in (b) (black lines for left panel and blue lines for right panel). (d) Critical curve for the emergence of type-III corner states.}\label{SSH}
\end{figure}

To study the band splittings, we consider three Su-Schrieffer-Heeger (SSH) models \cite{Ventra2022} as shown in Fig. \ref{SSH}(a), described by the following Hamiltonian $\mathcal{H}_i$ ($i=1,2,3$, we set $C_B=1$)
\begin{equation}
\mathcal{H}_1=\omega\left(
                \begin{array}{cc}
                  0 & C_N \\
                  C_N & 0 \\
                \end{array}
              \right),
\end{equation}

\begin{equation}
\mathcal{H}_2=\omega\left(
                \begin{array}{cccc}
                  0 & C_N & 0 & 0 \\
                  C_N & 0 & 1 & 0 \\
                  0 & 1 & 0 & C_N \\
                  0 & 0 & C_N & 0 \\
                \end{array}
              \right),
\end{equation}

\begin{equation}
\mathcal{H}_3=\omega\left(
                \begin{array}{cccccc}
                  0 & C_N & 0 & 0 & 0 & 0 \\
                  C_N & 0 & 1 & 0 & 0 & 0 \\
                  0 & 1 & 0 & C_N & 0 & 0 \\
                  0 & 0 & C_N & 0 & 1 & 0 \\
                  0 & 0 & 0 & 1 & 0 & C_N \\
                  0 & 0 & 0 & 0 & C_N & 0 \\
                \end{array}
              \right).
\end{equation}

The hopping terms $C_N$ cause the splitting of the boundary modes as
$\delta \omega_1=2\omega C_N,~\delta \omega_2=2\omega(C_N^2-C_N^4),~\delta \omega_3=2\omega(C_N^3-C_N^5)$, which can be approximatively regarded as an effective capacitances $C_N$, $C_N^2$ and $C_N^3$ to directly connect the two boundary nodes, as the red arrows shown in Fig. \ref{SSH}(a). Next, we will show how these connections induce the type-II and type-III corner states.

As shown in Fig. \ref{SSH}(b), one can separate the nodes of the breathing kagome lattice into three types: single, dimer, and trimer. The edge spectra are formed by the boundary dimers. Then, we reduce to the effective one-dimensional problem with two edges (green sectors) meeting at the corner, as shown in Fig. \ref{SSH}(b). The effective hopping term $\omega C_N$ connects the 3rd node to the 4th node, and as a result, the 1st and 6th bands split from the edge state continuum and form the type-II corner state [see black line in Fig. \ref{SSH}(c)]. To explain the energy splitting of the type-III corner state, we consider the interaction $\omega C_N^2$ between the 2nd and 5th, and the result is plotted in Fig. \ref{SSH}(c). We find the 2nd and 5th blue spectrums first submerge in the edge spectrums and then diverge from them at a critical point, and result in the emergence of the type-III corner states. Therefore, a threshold exists for the appearance of type-III corner states, as labeled by the orange dot in Fig. \ref{per}(d). In Fig. \ref{SSH}(d), we plot the critical points of emergence of type-III corner state for different capacitance $C_A$, and one can find the type-III corner state in the blue region. The critical points can be fitted by the following curve
\begin{equation}
\frac{C_A}{C_B}=0.0058\exp\left(21.11\frac{C_N}{C_B}\right),
\end{equation}
and different $C_A$ gives different threshold $C_{Nc}$. Furthermore, by taking the effective coupling $\omega C_N^3$ between 1st and 6th nodes into account, we find the 3rd and 4th can deviate from the edge state continuum a little but cannot escape from them. Therefore, one can only observe two types of corner states splitting from the edge spectrums in the present model.

\section{The realization of the negative resistance} \label{E}
As shown in Figs. \ref{Exp}(d) and \ref{Exp}(e), the negative resistance is realized by an integrated operational amplifier XL5532 and three resistances $R_a,R_b$, and $R_n$.

For an ideal operational amplitude, $i_+=0$, $i_-=0$, and $u_d=0$ (most modern amplifiers have large gains and input impedances, so the analysis is feasible in a real circuit). According to Ohm's law, we obtain
\begin{equation}
\begin{aligned}
&v_{\rm in}-v_{\rm o}=iR_n,\\
&v_{\rm o}=\frac{R_a+R_b}{R_b}v_{\rm in}.
\end{aligned}
\end{equation}
Combining these two equations, we have
\begin{equation}
\frac{v_{\rm in}}{i}=-\frac{R_a}{R_b}R_n.
\end{equation}
Here, we choose $R_a=R_b=R_n=4.99$ k$\Omega$, so the network is equivalent to a negative resistance due to $v_{\rm in}=i(-R)$.
\end{appendix}

\end{document}